\documentclass[%
 reprint,
superscriptaddress,
 amsmath,amssymb,
 aps
]{revtex4-2}

\usepackage{graphicx}
\usepackage{dcolumn}
\usepackage{bm}
\usepackage{xcolor}

\begin{document}

\preprint{APS/123-QED}

\title{\Large \textbf{Dependence of the electronic structure of the EuS/InAs interface on the bonding configuration}}

\author{Maituo Yu}
\affiliation{Department of Materials Science and Engineering, Carnegie Mellon University, Pittsburgh, PA 15213, USA}
\author{Saeed Moayedpour}
\affiliation{Department of Chemistry, Carnegie Mellon University, Pittsburgh, PA 15213, USA}
\author{Shuyang Yang}
\author{Derek Dardzinski}
\affiliation{Department of Materials Science and Engineering, Carnegie Mellon University, Pittsburgh, PA 15213, USA}
\author{Chunzhi Wu}
\affiliation{Department of Civil and Environmental Engineering, Carnegie Mellon University, Pittsburgh, PA 15213, USA}
\author{Vlad S. Pribiag}
\affiliation{School of Physics and Astronomy, University of Minnesota, Minneapolis, MN 55455, USA}
\author{Noa Marom}
 \email{Electronic mail: nmarom@andrew.cmu.edu}
\affiliation{Department of Materials Science and Engineering, Carnegie Mellon University, Pittsburgh, PA 15213, USA}
\affiliation{Department of Chemistry, Carnegie Mellon University, Pittsburgh, PA 15213, USA}
\affiliation{Department of Physics, Carnegie Mellon University, Pittsburgh, PA 15213, USA}

\date{\today}% It is always \today, today,

\begin{abstract}
Recently, the EuS/InAs interface has attracted attention for the possibility of inducing magnetic exchange correlations in a strong spin-orbit semiconductor, which could be useful for topological quantum devices. We use density functional theory (DFT) with a machine-learned Hubbard $U$ correction [npj Comput. Mater. 6, 180 (2020)] to elucidate the effect of the bonding configuration at the interface on the electronic structure. For all interface configurations considered here, we find that the EuS valence band maximum (VBM) lies below the InAs VBM. In addition, dispersed states emerge at the top of the InAs VBM at the interface, which do not exist in either material separately. These states are contributed mainly by the InAs layer adjacent to the interface. They are localized at the interface and may be attributed to charge transfer from the EuS to the InAs. The interface configuration affects the position of the EuS VBM with respect to the InAs VBM, as well as the dispersion of the interface state. For all interface configurations studied here, the induced magnetic moment in the InAs is small. This suggests that this interface, in its coherent form studied here, is not promising for inducing equilibrium magnetic properties in InAs.
\end{abstract}

\maketitle

\section{\label{sec:level1} Introduction}

Majorana zero modes may provide a pathway to the realization of fault tolerant quantum computing \cite{kitaev2001unpaired,kitaev2003fault,sarma2015majorana,lutchyn2018majorana,nayak2008non,freedman2003topological}. 
Several types of hybrid materials systems have been proposed to produce Majorana zero modes. These include the surface of a 3D topological insulator interfaced with an s-wave superconductor; a 2D semiconductor with strong orbit-coupling in proximity to an s-wave superconductor; and a 1D semiconductor with the Majorana quasi-particles appearing at the two ends of the 1D superconductor wire \cite{das2012zero,lee2019transport,fu2008superconducting,alicea2010majorana,lutchyn2018majorana,oreg2010helical}. To induce superconductivity at the interface, the semiconductor is in contact with a superconductor, such as Al, Pb, and Nb\cite{mayer2019superconducting,lutchyn2010majorana,gunel2012supercurrent,gusken2017mbe,paajaste2015pb}. One challenge that may be encountered in such platforms is that a large magnetic field is required to drive the system into the topological regime, which may be detrimental to the superconductivity \cite{deng2012anomalous, he2018magnetic,leijnse2012introduction,alicea2010majorana}. 
A possible solution is to add a ferromagnet (FM) in contact with the semiconductor to induce magnetism in the semiconductor internally via non-equilibrium spin injection \cite{yang2020spinTransport}, stray fields \cite{Jiang_2020stray} or the magnetic proximity effect, thus obviating the need for an external magnetic field \cite{sau2010non}.  

FM/semiconductor heterostructures have been widely used for spintronics \cite{adelmann2005spin,lou2007electrical,zhu2001room}. 
Typically, the FM is a metal, such as Fe, Ni, and Co \cite{alvarado1992observation,hanbicki2002efficient, fert2001conditions}.
However, a metallic FM may induce undesirable states in the gap of a semiconductor and can have detrimental effects by shunting away the current in a hybrid device. A ferromagnetic insulator (FMI) may be advantageous for Majorana devices if it provides proximity induced magnetism in the semiconductor without introducing such deleterious effects \cite{sau2010generic}. 
Europium containing FMIs have a particularly high magnetic moment per Eu atom and a large exchange coupling \cite{wei2016strong}. Several studies have reported proximity induced magnetism in various materials in contact with europium oxide and europium chalcogenides, including the EuS/Bi$_2$Se$_3$ interface \cite{katmis2016high}, phosphorene on EuO\cite{chen2017proximity}, MoS$_2$-EuS heterojunctions \cite{liang2017magnetic}, and EuS/Al heterostructures \cite{PhysRevLett.61.637, PhysRevLett.110.097001, PhysRevLett.106.247001, diesch2018creation}. A strong interfacial exchange field has been detected at the EuS/Pt interface \cite{EuS-Pt}. Moreover, proximity to EuS may break the time-reversal symmetry and open a gap in the interface state of topological insulators \cite{PhysRevB.91.195310, PhysRevB.98.081403, SnSe-EuS}.  
 
EuS has a rock salt structure with less than 1\% lattice mismatch to InAs, which makes it favorable for epitaxial growth.  Therefore, the EuS/InAs interface could be an ideal option for a hybrid FMI-semiconductor-superconductor structure if the proximity-induced magnetism in the InAs is sufficiently strong. Recently, an epitaxial EuS/InAs interface with a (001) orientation has been grown and characterized \cite{liu2019coherent}. Angle-resolved photoemission spectroscopy (ARPES) has shown that the Eu 4$f$ states, which form the top of the valence manifold of EuS, lie below the valence band maximum of InAs and that an interfacial quantum well state forms in the InAs. However, proximity-induced magnetism in the InAs could not be detected in neutron and X‐ray reflectivity measurements. Based on this, it was concluded in Ref. \cite{liu2019coherent} that the magnetic proximity effect at the EuS/InAs interface is weak and has a small influence on the overall magnetism in InAs. Subsequent work on a hybrid structure of InAs nanowires, Al, and EuS, in which the EuS was in contact with both the InAs and the Al has argued for the presence of magnetic exchange effects \cite{liu2020_strayAndExchange}. Very recently zero-bias peaks, which are a signature of topological superconductivity, have been reported in such a hybrid structure \cite{vaitiekenas2021zero}. However, these observations could be attributed to distinct mechanisms, such as the EuS inducing magnetism in the Al \cite{PhysRevLett.106.247001, PhysRevLett.61.637}, rather than the InAs or changing stray magnetic fields during the complex magnetic reversal \cite{yang2020spinTransport}.

 To elucidate the electronic structure of the EuS/InAs interface and help resolve the question of whether and to what extent proximity induced magnetism is achieved in the InAs, we conduct first principles simulations within density functional theory (DFT). A challenge for DFT simulations of the EuS/InAs interface is that commonly used semi-local exchange-correlation functionals, such as the generalized gradient approximation of Perdew, Burke, and Erzerhof (PBE) \cite{perdew1996generalized} provide a poor description of EuS and InAs, producing no band gap for both materials \cite{yu2020machine, ghosh2004electronic,wachter1972optical,massidda1990structural}. 
 This may be corrected by using more accurate hybrid functionals, which contain a fraction of exact exchange \cite{kim2009accurate,schlipf2013structural}.
 However, hybrid functionals are impractical for simulations of large interface models, owing to their high computational cost. Alternatively, within the DFT+$U$ approach \cite{anisimov1991band,dudarev1998electron}, a Hubbard $U$ correction may be added to a semi-local DFT functional, such as the PBE functional. Recently, we have developed a method of machine learning the optimal value of the effective Hubbard $U$ parameter by Bayesian optimization (BO) \cite{yu2020machine}. The PBE+U(BO) method has been demonstrated to provide a reliable description of bulk EuS and InAs \cite{yu2020machine}, as well as InAs surfaces \cite{yang2020electronic} and interfaces \cite{yang2021first}. Here, we use PBE+U(BO) to study the electronic structure of the EuS/InAs (001) interface. The bonding configuration at the interface may significantly affect the electronic and magnetic properties \cite{doi:10.1063/1.4858400}, therefore we consider different interface configurations. 
 
 We find that the interface configurations studied here share qualitatively similar electronic and magnetic properties. However, the quantitative details differ between configurations. In all interface configurations the valance band maximum (VBM) of the EuS lies below the VBM of the InAs and a dispersed interface state emerges in agreement with the ARPES experiments reported in Ref. \cite{liu2019coherent}. The interface configuration affects the extent of charge transfer from EuS to InAs, and as a result, the band alignment between the EuS and InAs, as well as the interface state dispersion. Greater charge transfer pushes the EuS VBM lower with respect to the InAs VBM and leads to more dispersed interface states. In all configurations studied here, the induced magnetic moment in the InAs is found to be small and short-ranged, which is also in agreement with Ref. \cite{liu2019coherent}. This indicates that the EuS/InAs interface, in its coherent form, is not particularly well suited for inducing magnetic exchange correlations in InAs.

\begin{figure}[h]
\centering
\includegraphics[width=0.5\textwidth]{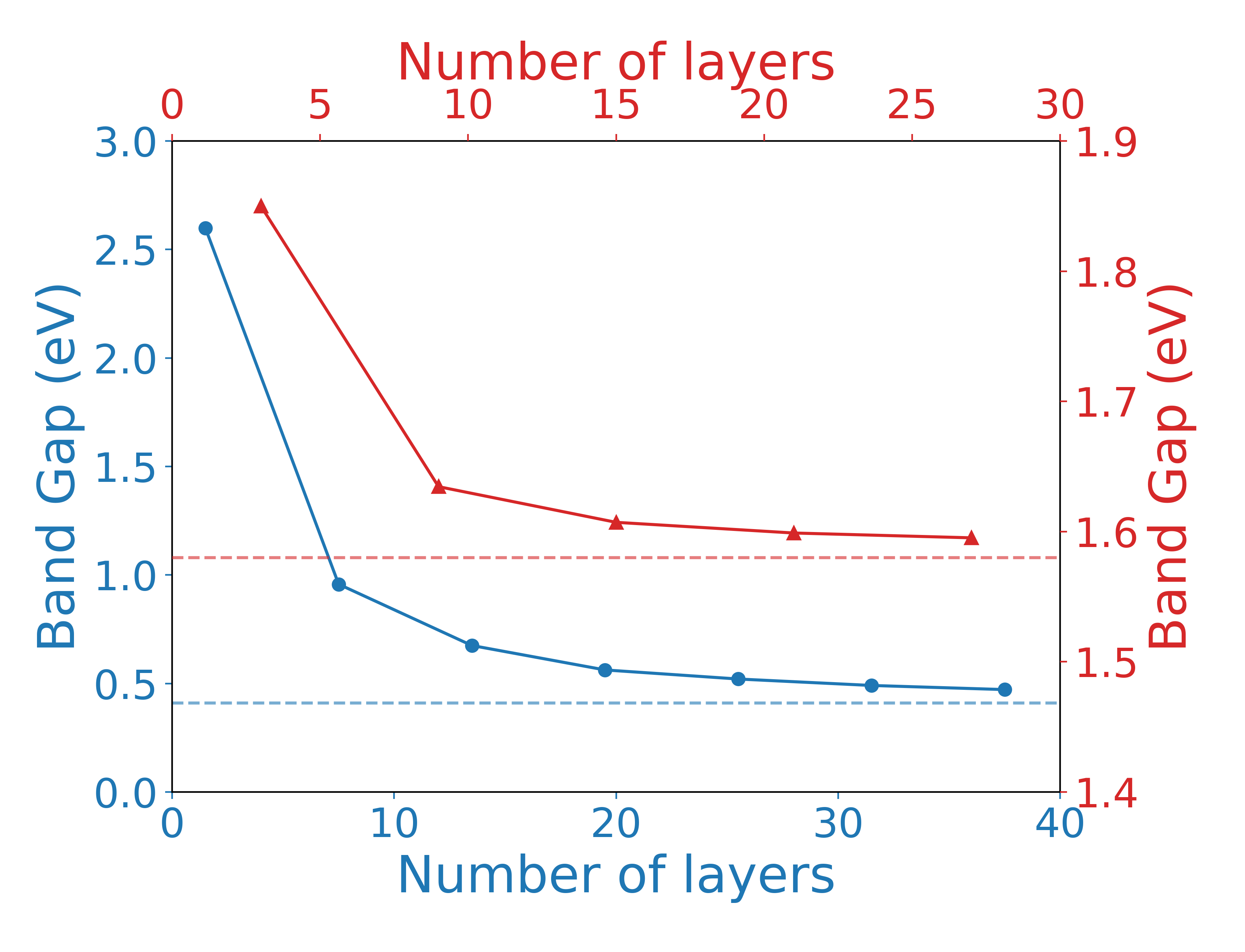}
\caption{\label{fig:slab_convergence} The band gap of (a) InAs and (b) EuS slabs as a function of the number of atomic layers. Dashed lines denote the bulk limit of 0.41 eV for InAs and 1.58 eV for EuS. }
\end{figure}

\begin{figure*}
\centering
\includegraphics[width=1\textwidth]{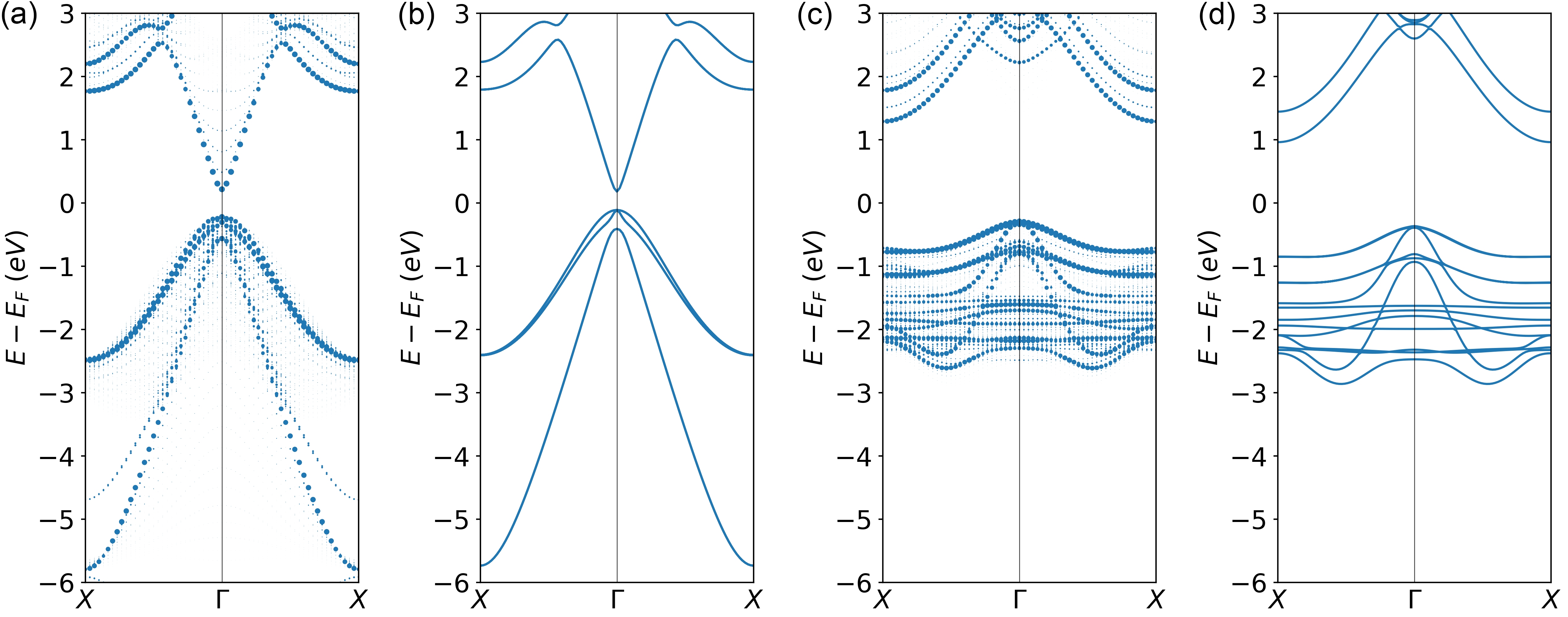}
\caption{\label{fig:slab_bands} (a) Bulk unfolded band structure of an InAs slab with 75 layers compared to (b) the band structure of bulk InAs oriented in the (001) direction.  (c) Bulk unfolded band structure of a EuS slab with 21 layers compared to (d) the band structure of bulk EuS oriented in the (001) direction.  }
\end{figure*}

\section{\label{sec:level2} Methods}

DFT calculations were performed using the Vienna \textit{ab initio} simulation package (VASP) with the projector augmented wave method (PAW) \cite{kresse1993ab, kresse1999ultrasoft, blochl1994projector}. PBE+U was used to describe the exchange-correlation interactions among electrons. The Hubbard $U$ correction was applied within the Dudarev approach \cite{dudarev1998electron} to the $p$ orbitals of In and As and the $f$ orbitals of Eu. The effective U parameters, $U_{eff} = U- J$, which correspond to the difference between the on-site Coulomb interaction, $U$, and the exchange interaction, $J$, were machine learned by Bayesian optimization (BO), as described in Ref. \cite{yu2020machine}. The $U_{eff}$ values obtained therein were $U_{eff}^{In,p}$= -0.5 eV, $U_{eff}^{As,p}$= -7.5 eV, and $U_{eff}^{Eu,f}$= 8.4 eV. The PBE+U(BO) method with these parameters has been used successfully for InAs surfaces \cite{yang2020electronic} and the InAs/GaSb interface \cite{yang2021first}. A plane-wave basis set was used with a kinetic energy cutoff of 450 eV. The Brillouin zone was sampled using the Monkhorst-Pack scheme with an $8\times8\times1$ k-point mesh. Spin-orbit coupling (SOC) \cite{steiner2016calculation} was included in all calculations with the $z$ spin quantization axis, as well as the dipole correction \cite{neugebauer1992adsorbate}. To avoid spurious interactions between periodic replicas in slab models, a sufficiently large vacuum region of $\sim$40 {\AA} was added in the $z$ direction. To eliminate surface states due to dangling bonds, the InAs slab was passivated by pseudo-hydrogen atoms with 1.25 electrons. In the structural relaxation and the interfacial distance optimization, the Tkatchenko-Scheffler (TS) pairwise dispersion method \cite{tkatchenko2009accurate} was used to describe the van der Waals interactions between the EuS film and the InAs substrate. The convergence criterion used in the structural relaxation was for the Hellman-Feynman
forces acting on ions to be below 0.01 eV/{\AA}. Bulk band unfolding \cite{yang2020electronic} was applied to all band structures to project the slab band structure onto the primitive cell and facilitate comparison with ARPES experiments. The spin-polarized density of states (DOS) of calculations with SOC was plotted by extracting the DOS of the majority spin channels and minority spin channels along the easy axis.

\section{\label{sec:level3} Results and discussion}

We begin by studying the band structures of InAs(001) and EuS(001) slabs separately.  Owing to the effect of quantum confinement, the band gap of a slab model decreases as the number of layers increases, eventually approaching the bulk band gap, as shown in Fig \ref{fig:slab_convergence}.   
The band gap of InAs converges with 75 atomic layers and the band gap of EuS converges with 21 atomic layers.
The bulk-unfolded band structures of these slab models of InAs and EuS are compared to bulk band structures with a (001) orientation in Fig. \ref{fig:slab_bands}. With this number of layers the band gap of InAs is 0.02 eV larger than the bulk limit and the band gap of EuS is 0.22 eV larger than the bulk limit. The main features of the bulk band structure are reproduced well by the slab models.

\begin{figure}[h]
\centering
\includegraphics[width=0.5\textwidth]{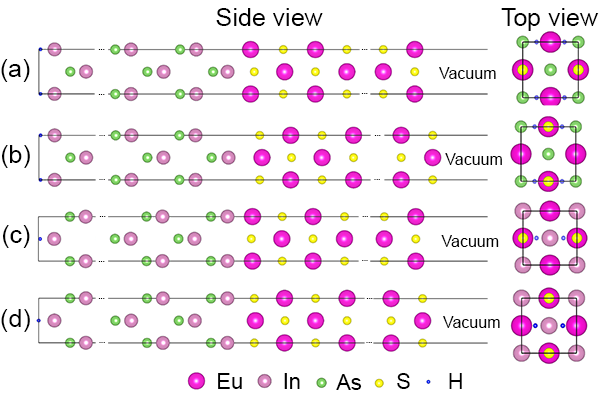}
\caption{\label{fig:configurations} Side view and top view of four configurations of an ideal EuS/InAs interface. From top to bottom, the configurations are labeled as (a) configuration 1 (C1), (b) configuration 2 (C2), (c) configuration 3 (C3), and (d) configuration 4 (C4).}
\end{figure}

To model the EuS/InAs interface, we assumed that an epitaxially matched EuS film would grow on top of In-terminated InAs(001), based on the experiments reported in Ref. \cite{liu2019coherent}. Therefore, the lattice constant of the InAs substrate was fixed at 6.06 {\AA} and the EuS film was strained to match it. Four unique interface configurations are possible with the EuS atoms positioned directly above the InAs lattice sites, as shown in Fig. \ref{fig:configurations}a-d. In Ref. \cite{liu2019coherent} configuration C1 (shown in Fig. \ref{fig:configurations}a) was proposed based on high-angle annular dark-field (HAADF) imaging
using an aberration-corrected scanning transmission electron
microscope (AC-STEM). It was also noted therein that single layer intermixing caused a displacement from ideal atomic rows and that atomic steps were likely present at the interface. Here, we only consider ideal interfaces. Intermixing, atomic steps, and the possible co-existence of different interface configurations are not taken into account.

\begin{figure}[h]
\centering
\includegraphics[width=0.5\textwidth]{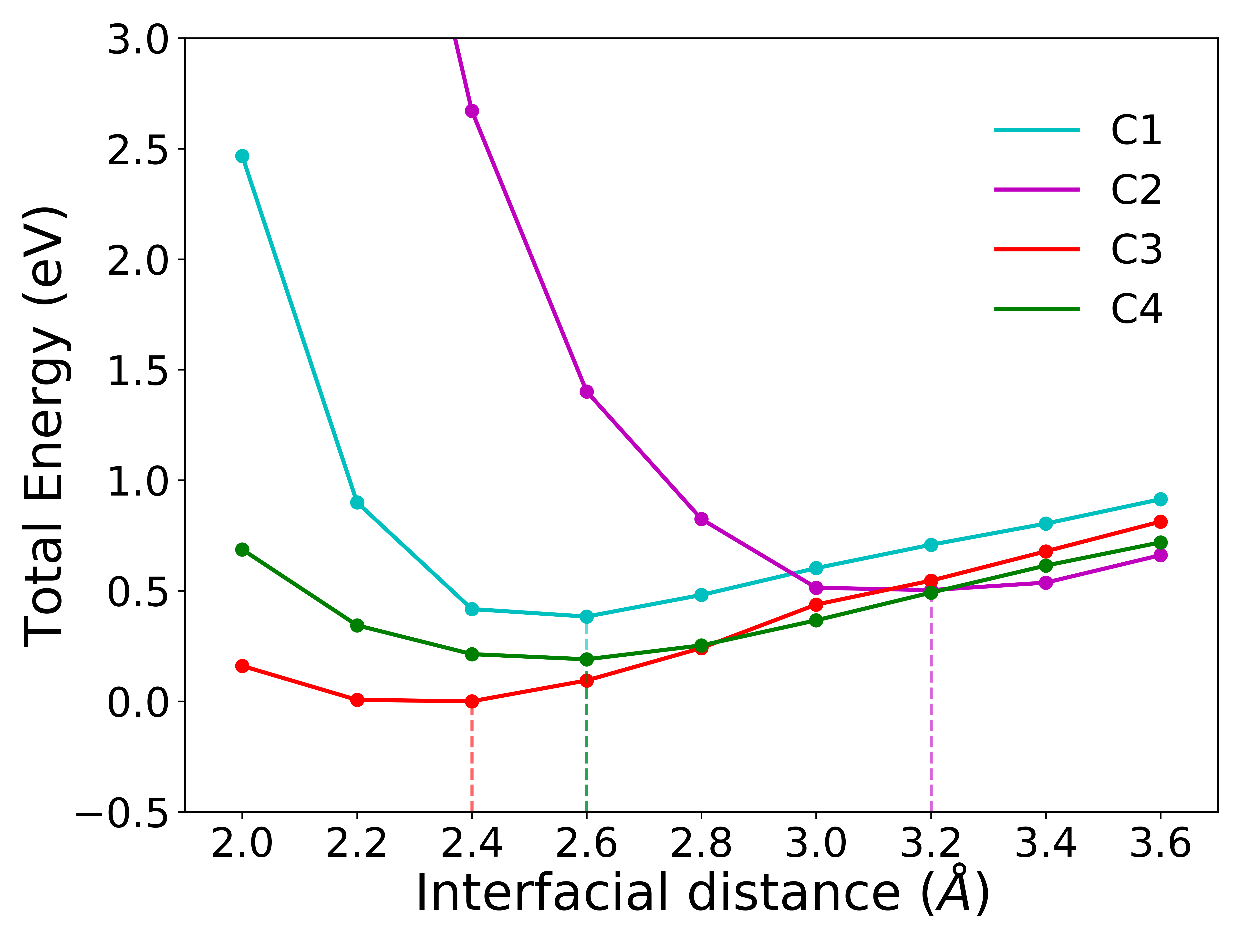}
\caption{\label{fig:interfacial_distance} Total energy change as a function of the interfacial distance for the four configurations. The lowest minimum is referenced to zero.  }
\end{figure}

To determine the interfacial distance to use as a starting point for relaxation for each configuration, the EuS film was moved along $z$ direction with respect to the InAs substrate with a step size of 0.1 {\AA}. The resulting energy curves are shown in Fig. \ref{fig:interfacial_distance}. We find that the interface configuration significantly affects the interfacial distance. The largest interfacial distance of 3.2 {\AA} is found for configuration C2 and the smallest interfacial distance of 2.4 {\AA} is found for configuration C3.

\begin{figure}[h]
\centering
\includegraphics[width=0.5\textwidth]{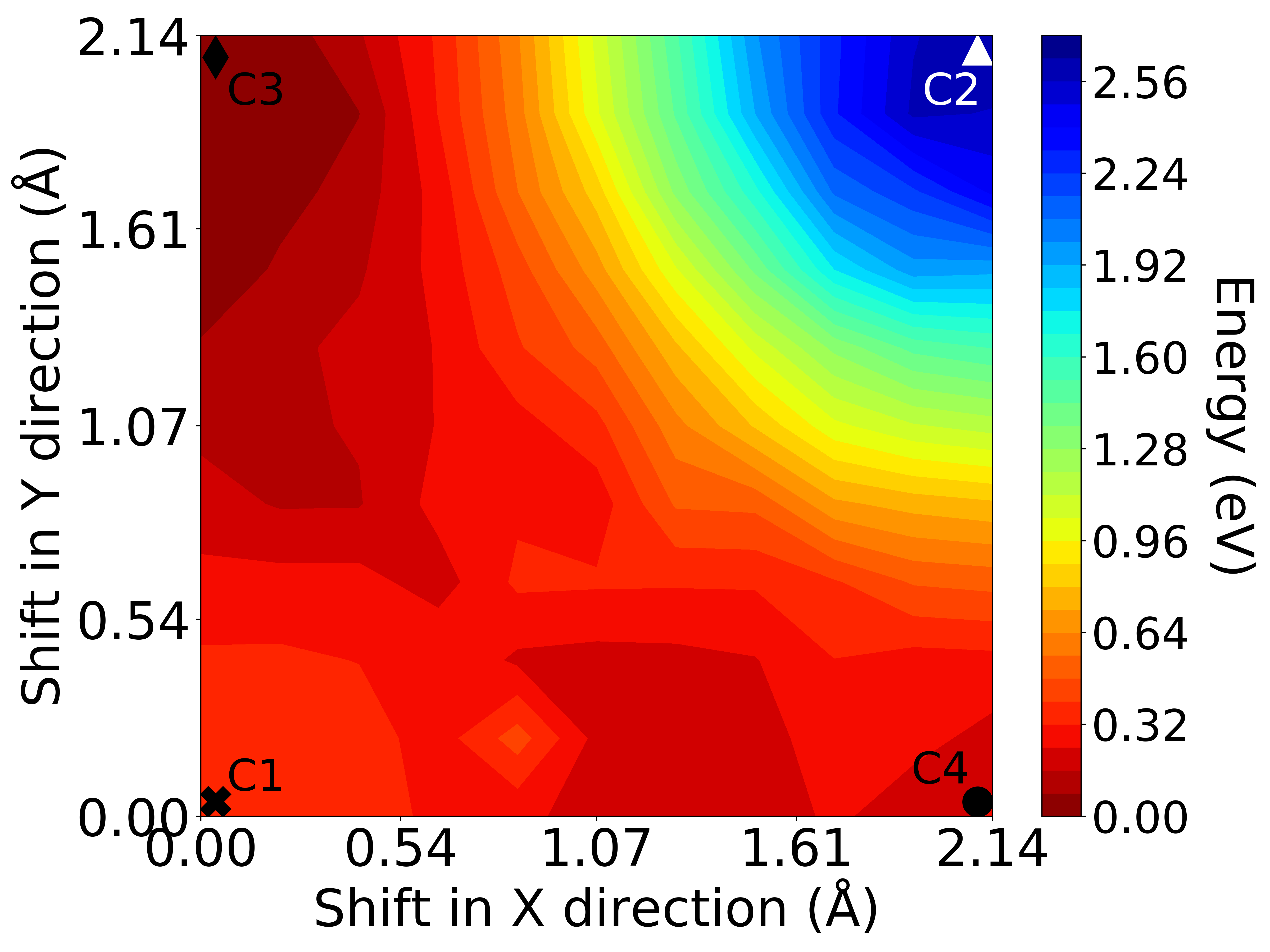}
\caption{\label{fig:PES} PES scan at the interfacial distance of 2.4 $\AA$}
\end{figure}

A potential energy surface (PES) scan was performed by shifting the EuS film in the $xy$ plane at a fixed distance of 2.4 {\AA} from the InAs substrate along $z$. The PES scan was performed with 11 layers of InAs and 3 layers of EuS. We performed a convergence test with a varying number of layers of InAs and EuS to verify that the energy ranking exhibited in the PES is not affected by the number of layers.  The resulting PES is shown in Fig. \ref{fig:PES}. Configuration C3 is the global minimum. Configuration C4 is a local minimum. Configuration C2 is the global maximum. Based on our calculations, configuration C1, reported in Ref. \cite{liu2019coherent} is, in fact, a local maximum. 

\begin{figure}[h]
\centering
\includegraphics[width=0.5\textwidth]{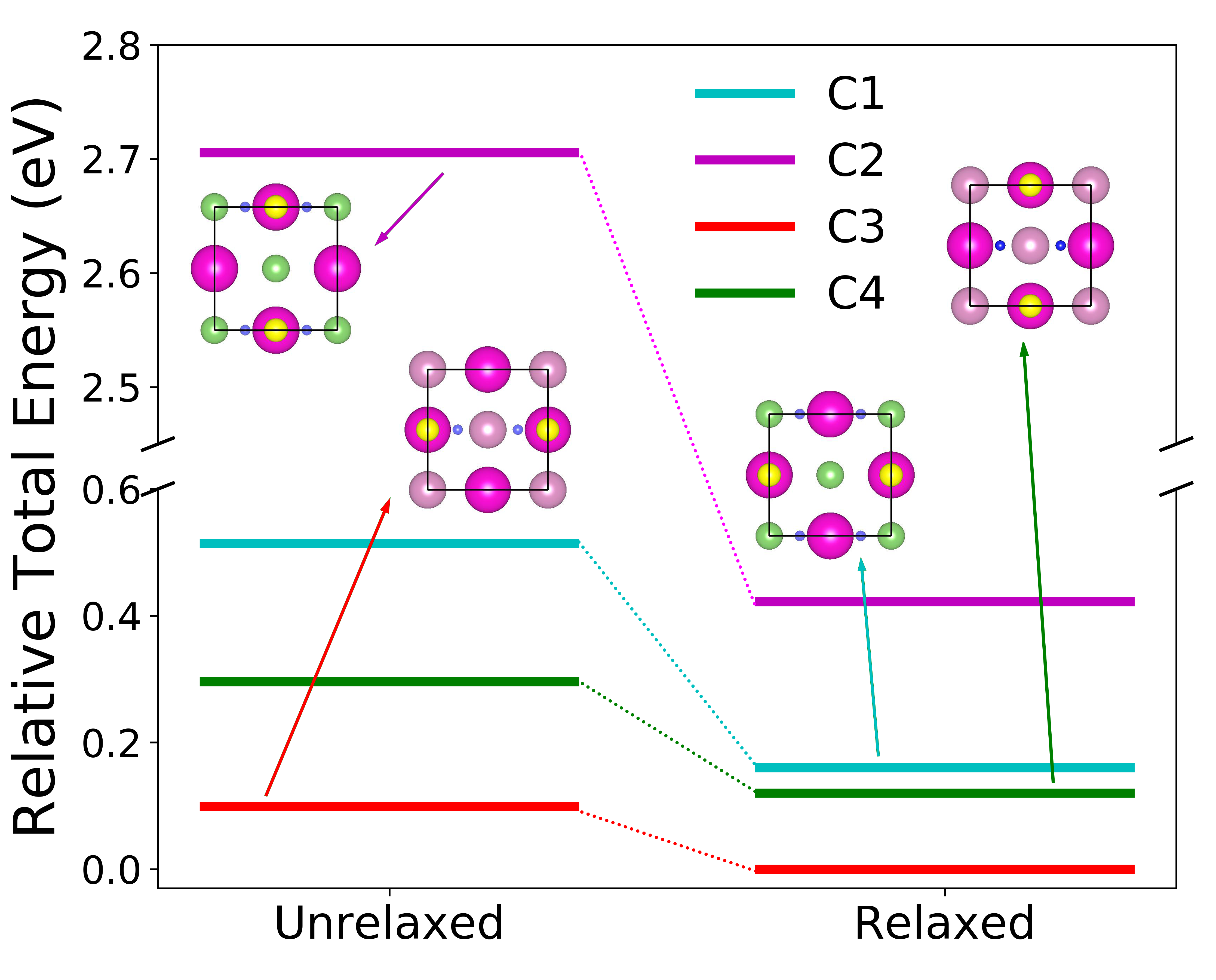}
\caption{\label{fig:ranking} Energy ranking of the four interface configurations before and after relaxation. The lowest energy structure after relaxation is referenced to zero.}
\end{figure}

Starting from the optimal interfacial distance obtained for each configuration, geometry relaxation was performed for the first 6 InAs layers and 2 EuS layers around the interface. The atomic positions were constrained in the $xy$ plane and allowed to vary in the $z$ direction. 
%For configuration C1, geometry relaxation reduced the interfacial distance from 2.60 $\AA$ to 2.56 $\AA$ and the interatomic distance between Eu and In at the interface decreased from 3.99 $\AA$ to 3.75 $\AA$. For configuration C2, the closest distance between the In and S atoms at the interface changed the most during relaxation, from 4.41 $\AA$ to 4.27 $\AA$. For configurations C3 and C4, relaxation led to compression of the first two EuS layers at the interface and a stretch of the InAs layers, such that overall the interatomic distance of In-Eu and In-S did not change much.  
Fig. \ref{fig:ranking} shows the relative energy ranking of the four configurations before and after relaxation. Relaxation resulted in reduction of the energies of all structures, however it did not affect the stability ranking. After relaxation, configuration C3 is still the most stable, followed by configurations C4, C1, and C2. Configuration C2 is the least stable owing to unfavorable cation-cation bonding at the interface. The higher stability of configurations C3 and C4 may be attributed to the tetrahedral site being favored over the edge-center site because it matches better with the zincblende packing of InAs. The HAADF-STEM micrograph of the EuS/InAs interface shown in Ref. \cite{liu2019coherent} could also be compatible with the C3 model in terms of atomic positions. However, a more complex 3D structure including, for example, surface atomic steps, would be needed to explain the HAADF intensity reduction and atomic column elongation in the positions nearby the interface for a full matching of the experimental image with the C3 epitaxial arrangement.

\begin{figure*}
\centering
\includegraphics[width=0.9\textwidth]{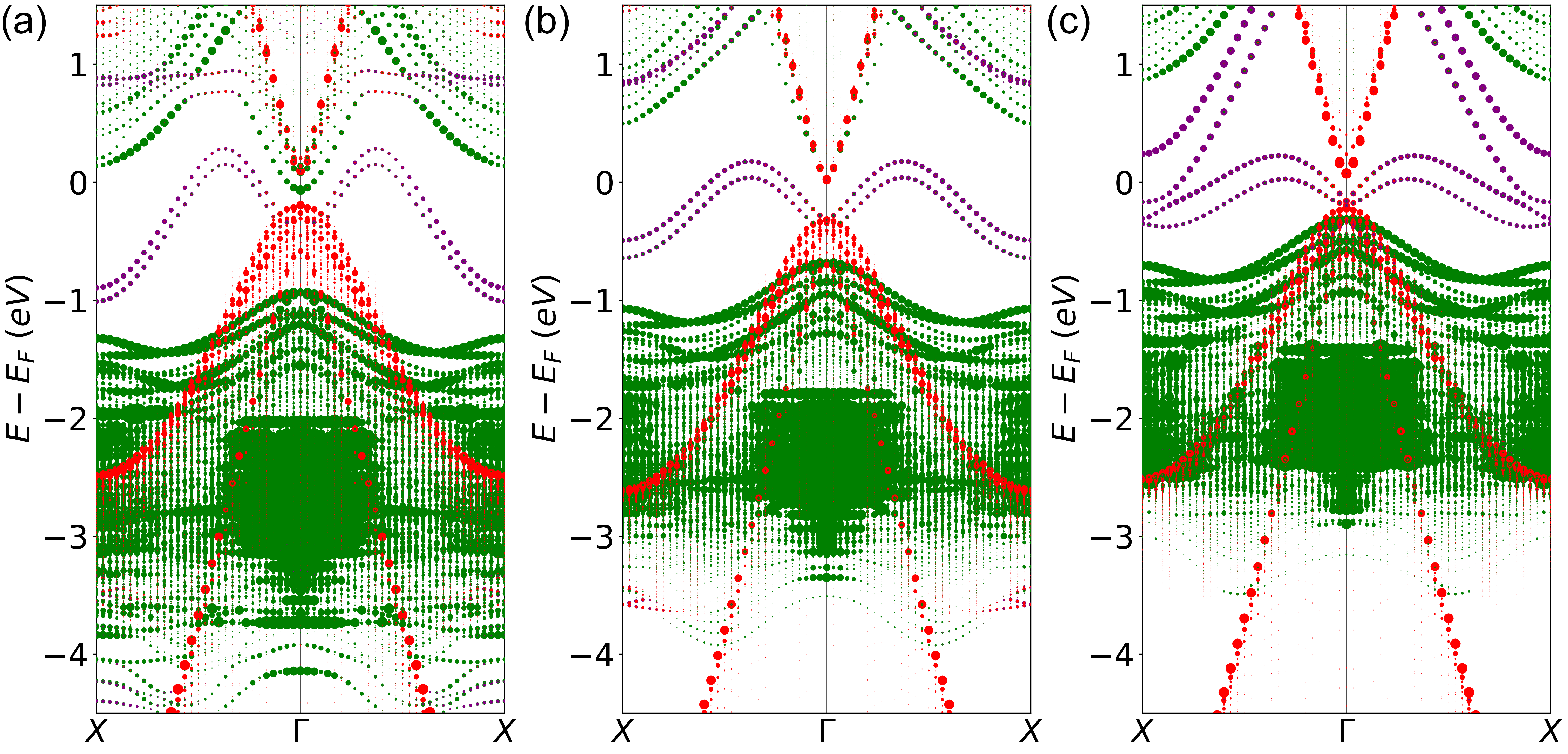}
\caption{\label{fig:interface_bands} Unfolded band structure of different interface  configurations:(a) C1, (b) C3, and (c) C4. The red, green, and purple markers denote the contributions from InAs, EuS, and the interface states, respectively. }
\end{figure*}

We now proceed to examine the effect of the atomistic configuration of the interface on its electronic structure. Because configuration C2 is unstable we do not consider it further. Configuration C1 is considered despite being a local maximum because it was proposed in Ref. \cite{liu2019coherent}. Fig. \ref{fig:interface_bands} shows bulk-unfolded band structures of configurations C1, C3, and C4. For all three configurations the the top of the EuS valence band (green) is found below the top of the InAs valence band (red) at the $\Gamma$ point in agreement with the ARPES results from Ref. \cite{liu2019coherent}. However, the interface configuration significantly affects the band alignment: for configuration C1, the EuS VBM is found 0.72 eV below the InAs VBM; for configuration C3, the EuS VBM is found 0.36 eV below the InAs VBM; and for configuration C4 it is found 0.09 eV below the InAs VBM. The band alignment produced by configuration C3 appears to be in better agreement with the ARPES results from Ref. \cite{liu2019coherent} than configuration C1.

\begin{figure}[h]
\centering
\includegraphics[width=0.42\textwidth]{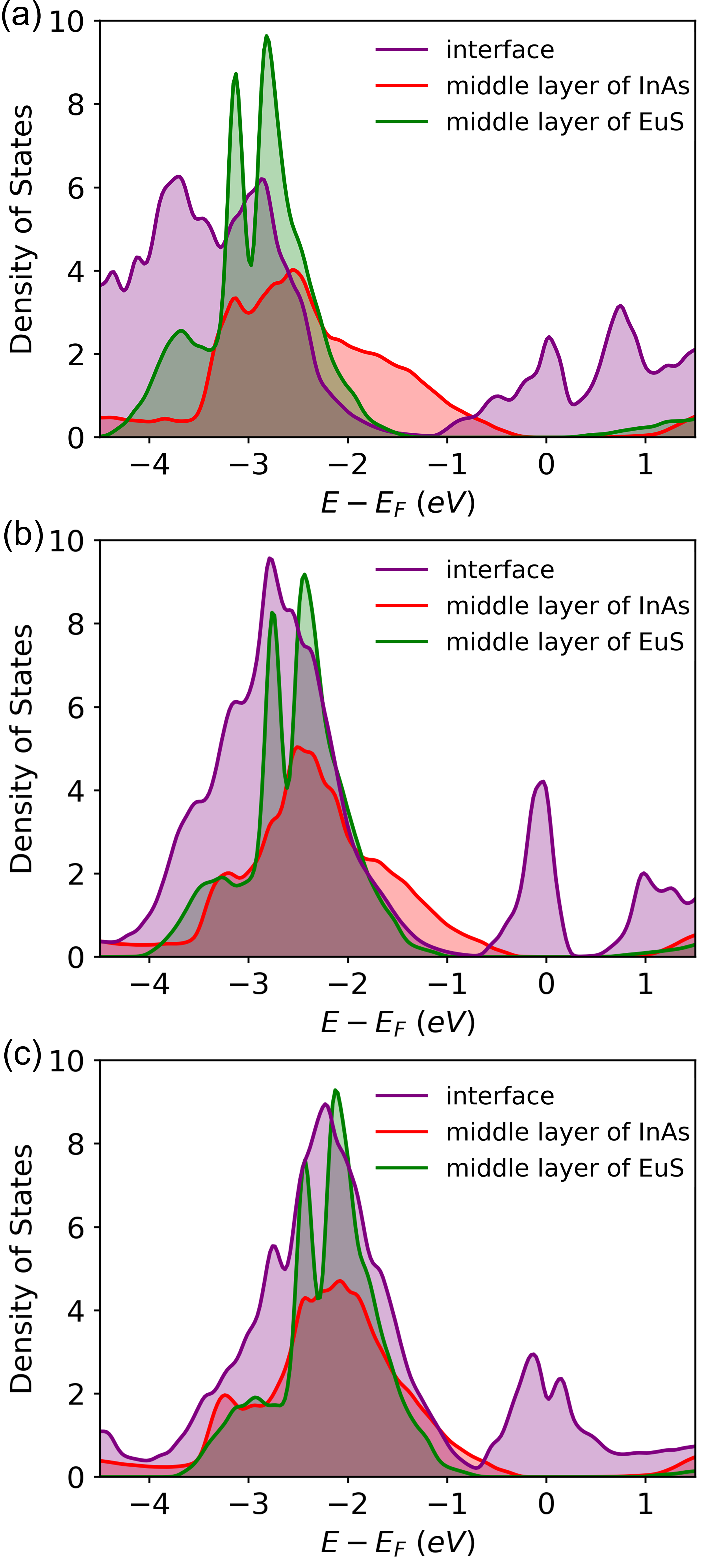}
\caption{\label{fig:DOS} Local density of states (LDOS) of (a) configuration C1, (b) configuration C3, and (c) configuration C4. The interface layer refers to the first InAs and EuS layers at the interface; The middle layer of InAs refers to the 38$^{th}$ InAs layer from the interface; The middle layer of EuS refers to the 11$^{th}$ EuS layer from the interface. The Fermi levels are shifted to 0 eV.}
\end{figure}

\begin{figure}[h]
\centering
\includegraphics[width=0.5\textwidth]{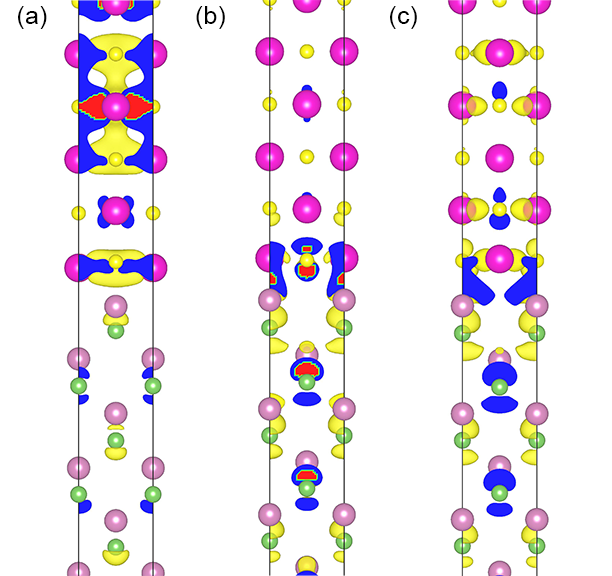}
\caption{\label{fig:WF} The partial charges associated
with the interface states of (a) configuration C1, (b) configuration C3, and (c) configuration C4 at $\Gamma$ point. The electron charge isosurface is colored in yellow  and its intersections with the unit cell boundary are colored in blue. Hollow spaces in the isosurface that intersect with the unit cell boundary are colored in red.}
\end{figure}

Interestingly, dispersed states, colored in purple in Fig. \ref{fig:interface_bands}, appear at the top of the valence band at the EuS/InAs interface, which are not present in either InAs or EuS alone (see Fig. \ref{fig:slab_bands}). By resolving the atomic contributions to the band structure, we find that these interface states are contributed predominantly by the first InAs layer at the interface. The interface states found here are in agreement with the quantum well state observed in ARPES in Ref. \cite{liu2019coherent}. The dispersion of the interface states depends on the interface configuration with configuration C1 producing the greatest band dispersion and configuration C4 yielding the smallest band dispersion. The local density of states (LDOS), shown in Fig. \ref{fig:DOS}, shows that the interface states (purple) are present around the Fermi level only at the interface. Away from the interface, the bulk DOS of InAs (red) and EuS (green) is recovered. Fig. \ref{fig:WF} shows the partial charges associated with the interface states at the $\Gamma$ point. The interface configuration also affects the wave-function of the interface state: For configuration C1 the interface state wave-function is predominantly localized in the EuS side of the interface; for configuration C3 the interface state wave-function is predominantly localized in the InAs side of the interface; and for configuration C4 the interface state wave-function is evenly distributed across the InAs and EuS. 

\begin{figure}[h]
\centering
\includegraphics[width=0.5\textwidth]{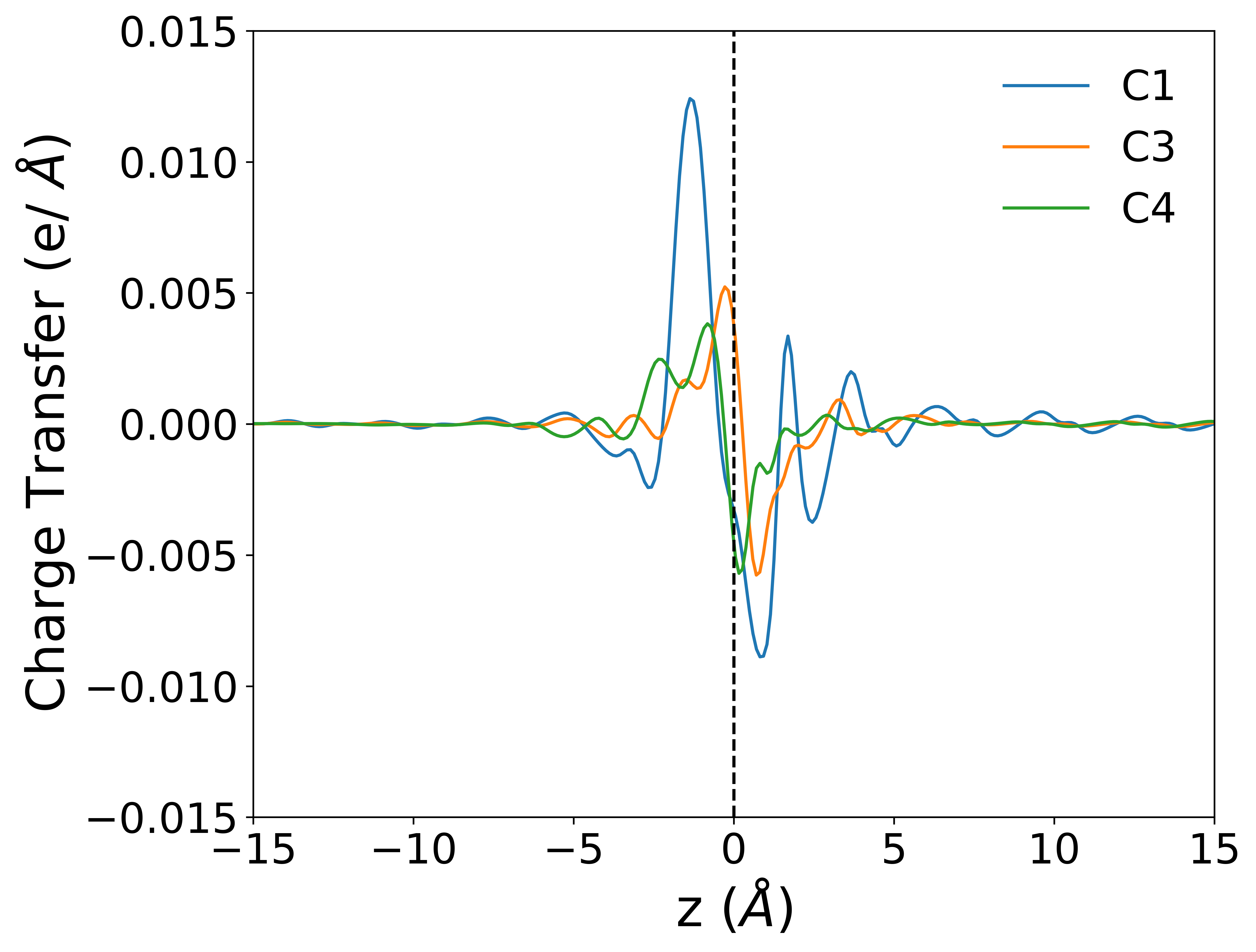}
\caption{\label{fig:CT} Charge transfer as a function of distance from the interface for InAs/EuS with different configurations. The center of the interface is referenced to 0. The InAs is on the left side and the EuS is on the right side.}
\end{figure}

The effect of the interface configuration on the band alignment and the dispersion of the interface state may be explained by the extent of charge transfer at the interface. The net charge transfer was calculated by:
\begin{equation}
C_{net}(z) = C[interface] - C[substrate] - C[film]
\end{equation}
where $C[\cdot]$ is the charge averaged over the $xy$ plane along the $z$-axis \cite{SnSe-EuS}.  The center of the interface is defined as $z = 0$. To evaluate the charge transfer, DFT calculations were performed for the interface slab and for separate slabs containing only the substrate and only the film with the same geometry as the interface slab. Fig. \ref{fig:CT} shows the resulting charge transfer for different interface configurations. $C_{net}$ is positive at the interfacial layer of InAs and negative at the first layer of EuS, meaning that charge is transferred from the EuS to the InAs. Configuration C1 exhibits the greatest charge transfer, followed by configuration C3 and configuration C4. The larger the charge transfer, the lower the EuS VBM lies below the InAs VBM and the greater the band dispersion of the interface state.

\begin{figure}[h]
\centering
\includegraphics[width=0.5\textwidth]{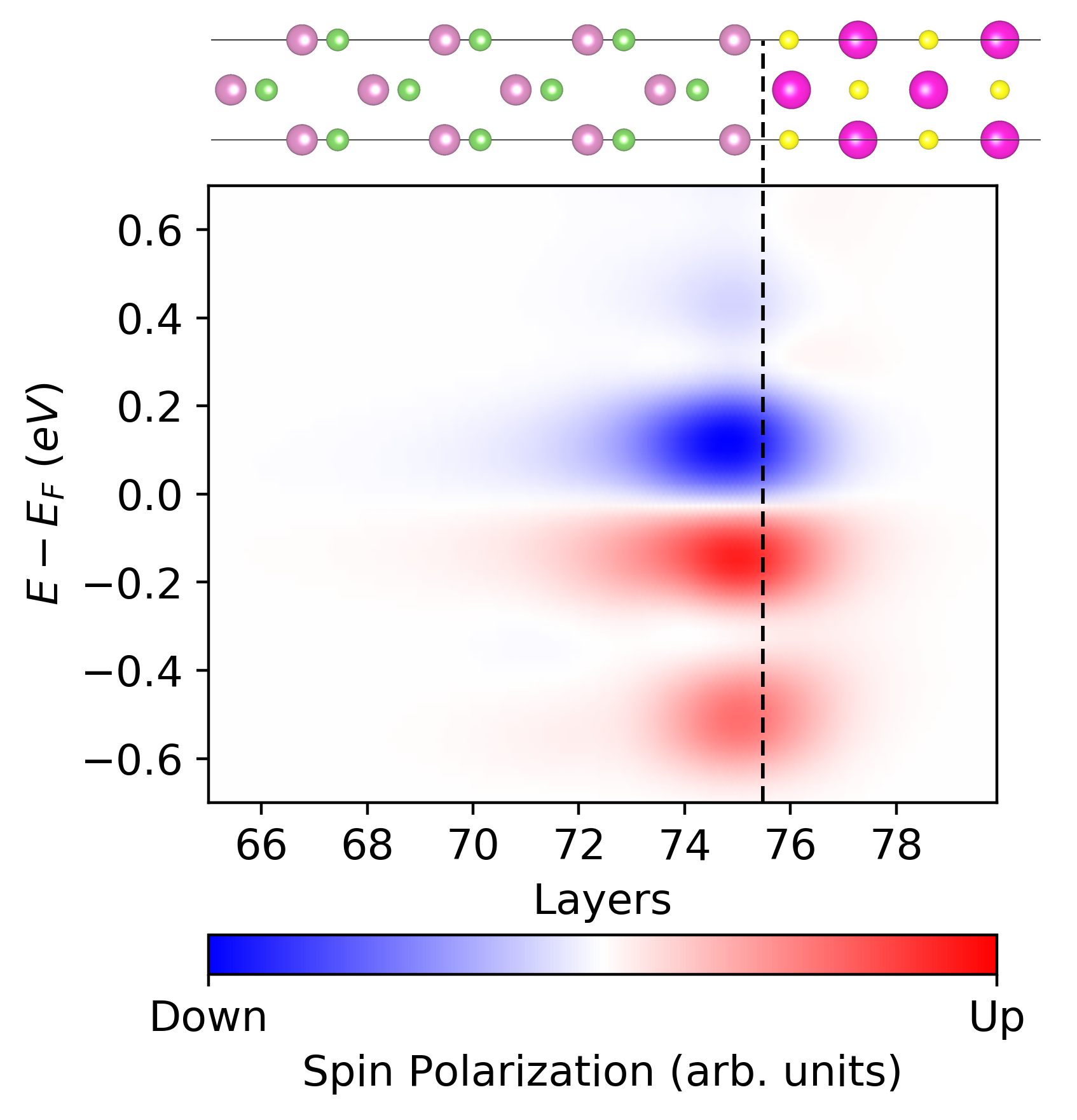}
\caption{\label{fig:SPDOS} Spin polarization (the difference between the majority DOS and the minority DOS) as a function of the layer index for configuration C3. Up and down refer to the spin-majority DOS and spin-minority DOS. The position of the interface is indicated by the black dashed line.}
\end{figure}

The computed magnetic moment of the Eu atoms in bulk EuS is found to be 7.0 $\mu$B, as expected. However, the proximity-induced magnetic moment in the InAs is small for all three interface configurations. The magnetic moment in the first layer of InAs (the In atom) adjacent to the EuS is 0.013 $\mu$B  for configuration C1, 0.087 $\mu$B for configuration C3, and 0.166 $\mu$B for configuration C4. The magnetic moment decays rapidly with the distance from the interface and vanishes completely by the 8$^{th}$ atomic layer of InAs from the interface. The induced magnetic moment is inversely correlated with the extent of charge transfer at the interface. In Ref. \cite{liu2019coherent}, a DFT calculation was conducted for a small interface model with two unit cells of each material in configuration C1  and 10 $\AA$ of vacuum without surface passivation. PBE+U was used with a $U$ value of 6 eV for the $f$ orbital of Eu. They found a magnetic moment of 0.07 $\mu$B on the In atom adjacent to the EuS. Despite the small induced magnetic moment, the DOS at the interface is spin-polarized, as shown in Fig. \ref{fig:SPDOS}. About 0.25 eV below the Fermi level the DOS is dominated by the majority spin channel, whereas the minority spin channel dominates about 0.2 eV above the Fermi level. This spin-polarization is highly localized to the first few InAs layers near the interface - the same region where the dispersed interface states exist. 
Based on our results, it is unlikely that the proximity-induced magnetic moment in InAs is sufficiently strong to account for large effects such as the recent observations of magnetic hysteresis and of zero-bias conductance peaks at zero magnetic field in EuS/InAs/Al hybrid quantum devices \cite{liu2020_strayAndExchange, vaitiekenas2021zero}. However, weakly spin-polarized in-plane transport may be possible via the interface state.

\section{\label{sec:level4} Conclusion}

In summary, we have conducted a first-principles investigation of the effect of the bonding configuration on the electronic properties of the EuS/InAs (001) interface, using DFT with a machine-learned Hubbard U correction. The DFT+U(BO) method enables unprecedented simulations of large, well-converged interface models. For all interface configurations studied here, the valence band of the EuS lies below the valence band of the InAs. In addition, a dispersed interface state appears at the top of the InAs valence band, which is contributed primarily by the InAs layer closest to the interface. These findings are in agreement with ARPES experiments reported in Ref. \cite{liu2019coherent}. The different interface configurations differ in the band alignment between the EuS and InAs and in the dispersion of the interface state. Both are correlated with the extent of charge transfer from EuS to InAs at the interface. The greater the charge transfer the lower the EuS VBM lies below the InAs VBM and the more dispersed the interface state becomes. For all interface configurations studied here, the induced magnetic moment in the InAs is found to be modest and short-ranged, which is also in agreement with Ref.  \cite{liu2019coherent}. This suggests that EuS/InAs coherent interfaces are not ideally suited for applications which seek to leverage equilibrium magnetic moments induced in a III-V semiconductor by proximity. This work demonstrates that first principles simulations may help interpret experimental findings and elucidate the electronic properties as they relate to the interface structure.

\section{\label{sec:level5} Acknowledgments}

 We thank Peter Krogstrup and Yu Liu from University of Copenhagen and Microsoft Quantum Materials Lab Copenhagen and Sara Martí-Sanchez from the Catalan Institute of Nanoscience and Nanotechnology for helpful discussions of the details of their HAADF-STEM experiments and DFT calculations. We thank Sergey Frolov from the University of Pittsburgh, Paul Crowell from the University of Minnesota, and Chris Palmstr{\o}m from the University of California Santa Barbara for helpful discussions. This research was funded by the Department of Energy through grant DE-SC0019274. This research used resources of the National Energy Research Scientific Computing Center (NERSC), a DOE Office of Science User Facility supported by the Office of Science of the U.S. Department of Energy under contract no. DE-AC02-05CH11231.

% \bibliography{reference} %your .bib file

\begin{thebibliography}{10}

\bibitem{kitaev2001unpaired}
A.~Y. Kitaev.
\newblock Unpaired majorana fermions in quantum wires.
\newblock {\em Phys.-Uspekhi}, 44(10S):131, 2001.

\bibitem{kitaev2003fault}
A.~Y. Kitaev.
\newblock Fault-tolerant quantum computation by anyons.
\newblock {\em Ann. Phys. (N.Y.)}, 303(1):2--30, 2003.

\bibitem{sarma2015majorana}
S.~D. Sarma, M.~Freedman, and C.~Nayak.
\newblock Majorana zero modes and topological quantum computation.
\newblock {\em NPJ Quantum Inf.}, 1(1):1--13, 2015.

\bibitem{lutchyn2018majorana}
R.~M. Lutchyn, E.~P. A.~M. Bakkers, L.~P. Kouwenhoven, P.~Krogstrup, C.~M.
  Marcus, and Y.~Oreg.
\newblock Majorana zero modes in superconductor--semiconductor
  heterostructures.
\newblock {\em Nat. Rev. Mater.}, 3(5):52--68, 2018.

\bibitem{nayak2008non}
C.~Nayak, S.~H. Simon, A.~Stern, M.~Freedman, and S.~D. Sarma.
\newblock Non-abelian anyons and topological quantum computation.
\newblock {\em Rev. Mod. Phys.}, 80(3):1083, 2008.

\bibitem{freedman2003topological}
M.~Freedman, A.~Kitaev, M.~Larsen, and Z.~H. Wang.
\newblock Topological quantum computation.
\newblock {\em Bull. Am. Math. Soc.}, 40(1):31--38, 2003.

\bibitem{das2012zero}
A.~Das, Y.~Ronen, Y.~Most, Y.~Oreg, M.~Heiblum, and H.~Shtrikman.
\newblock {Zero-bias peaks and splitting in an Al--InAs nanowire topological
  superconductor as a signature of Majorana fermions}.
\newblock {\em Nat. Phys.}, 8(12):887--895, 2012.

\bibitem{lee2019transport}
J.~S. Lee, B.~Shojaei, M.~Pendharkar, A.~P. McFadden, Y.~Kim, H.~J. Suominen,
  M.~Kjaergaard, F.~Nichele, H.~Zhang, and C.~M. Marcus.
\newblock {Transport Studies of Epi-Al/InAs Two-Dimensional Electron Gas
  Systems for Required Building-Blocks in Topological Superconductor Networks}.
\newblock {\em Nano Lett.}, 19(5):3083--3090, 2019.

\bibitem{fu2008superconducting}
L.~Fu and C.~L. Kane.
\newblock Superconducting proximity effect and majorana fermions at the surface
  of a topological insulator.
\newblock {\em Phys. Rev. Lett.}, 100(9):096407, 2008.

\bibitem{alicea2010majorana}
J.~Alicea.
\newblock Majorana fermions in a tunable semiconductor device.
\newblock {\em Phys. Rev. B}, 81(12):125318, 2010.

\bibitem{oreg2010helical}
Y.~Oreg, G.~Refael, and F.~Von~Oppen.
\newblock Helical liquids and majorana bound states in quantum wires.
\newblock {\em Phys. Rev. Lett.}, 105(17):177002, 2010.

\bibitem{mayer2019superconducting}
W.~Mayer, J.~Yuan, K.~S. Wickramasinghe, T.~Nguyen, M.~C. Dartiailh, and
  J.~Shabani.
\newblock {Superconducting proximity effect in epitaxial Al-InAs
  heterostructures}.
\newblock {\em Appl. Phys. Lett.}, 114(10):103104, 2019.

\bibitem{lutchyn2010majorana}
R.~M. Lutchyn, J.~D. Sau, and S.~D. Sarma.
\newblock Majorana fermions and a topological phase transition in
  semiconductor-superconductor heterostructures.
\newblock {\em Phys. Rev. Lett.}, 105(7):077001, 2010.

\bibitem{gunel2012supercurrent}
H.~Y. G{\"u}nel, I.~E. Batov, H.~Hardtdegen, K.~Sladek, A.~Winden, K.~Weis,
  G.~Panaitov, D.~Gr{\"u}tzmacher, and T.~Sch{\"a}pers.
\newblock {Supercurrent in Nb/inas-nanowire/Nb josephson junctions}.
\newblock {\em J. Appl. Phys.}, 112(3):034316, 2012.

\bibitem{gusken2017mbe}
N.~A. G{\"u}sken, T.~Rieger, P.~Zellekens, B.~Bennemann, E.~Neumann, M.~I.
  Lepsa, T.~Sch{\"a}pers, and D.~Gr{\"u}tzmacher.
\newblock {MBE growth of Al/InAs and Nb/InAs superconducting hybrid nanowire
  structures}.
\newblock {\em Nanoscale}, 9(43):16735--16741, 2017.

\bibitem{paajaste2015pb}
J.~Paajaste, M.~Amado, S.~Roddaro, F.~S. Bergeret, D.~Ercolani, L.~Sorba, and
  F.~Giazotto.
\newblock {Pb/InAs nanowire Josephson junction with high critical current and
  magnetic flux focusing}.
\newblock {\em Nano Lett.}, 15(3):1803--1808, 2015.

\bibitem{deng2012anomalous}
M.~T. Deng, C.~L. Yu, G.~Y. Huang, M.~Larsson, P.~Caroff, and H.~Q. Xu.
\newblock {Anomalous zero-bias conductance peak in a Nb--InSb nanowire--Nb
  hybrid device}.
\newblock {\em Nano Lett.}, 12(12):6414--6419, 2012.

\bibitem{he2018magnetic}
W.~Y. He, B.~T. Zhou, J.~J. He, N.~F.~Q. Yuan, T.~Zhang, and K.~T. Law.
\newblock Magnetic field driven nodal topological superconductivity in
  monolayer transition metal dichalcogenides.
\newblock {\em Commun. Phys.}, 1(1):1--7, 2018.

\bibitem{leijnse2012introduction}
M.~Leijnse and K.~Flensberg.
\newblock Introduction to topological superconductivity and majorana fermions.
\newblock {\em Semicond. Sci. Technol.}, 27(12):124003, 2012.

\bibitem{yang2020spinTransport}
Z.~D. Yang, B.~Heischmidt, S.~Gazibegovic, G.~Badawy, D.~Car, P.~A. Crowell,
  E.~P. A.~M. Bakkers, and V.~S. Pribiag.
\newblock {Spin Transport in Ferromagnet-InSb Nanowire Quantum Devices}.
\newblock {\em Nano Lett.}, 20(5):3232--3239, 2020.
\newblock PMID: 32338518.

\bibitem{Jiang_2020stray}
Y.~Jiang, E.~J. De~Jong, V.~van~de Sande, S.~Gazibegovic, G.~Badawy, E.~P.
  A.~M. Bakkers, and S.~M. Frolov.
\newblock Hysteretic magnetoresistance in nanowire devices due to stray fields
  induced by micromagnets.
\newblock {\em Nanotechnology}, 32(9):095001, dec 2020.

\bibitem{sau2010non}
J.~D. Sau, S.~Tewari, R.~M. Lutchyn, T.~D. Stanescu, and S.~D. Sarma.
\newblock Non-abelian quantum order in spin-orbit-coupled semiconductors:
  Search for topological majorana particles in solid-state systems.
\newblock {\em Phys. Rev. B}, 82(21):214509, 2010.

\bibitem{adelmann2005spin}
C.~Adelmann, X.~Lou, J.~Strand, C.~J. Palmstr{\o}m, and P.~A. Crowell.
\newblock Spin injection and relaxation in ferromagnet-semiconductor
  heterostructures.
\newblock {\em Phys. Rev. B}, 71(12):121301, 2005.

\bibitem{lou2007electrical}
X.~H. Lou, C.~Adelmann, S.~A. Crooker, E.~S. Garlid, J.~J. Zhang, K.~S.~M.
  Reddy, S.~D. Flexner, C.~J. Palmstr{\o}m, and P.~A. Crowell.
\newblock Electrical detection of spin transport in lateral
  ferromagnet--semiconductor devices.
\newblock {\em Nat. Phys.}, 3(3):197--202, 2007.

\bibitem{zhu2001room}
H.~J. Zhu, M.~Ramsteiner, H.~Kostial, M.~Wassermeier, H.~P. Sch{\"o}nherr, and
  K.~H. Ploog.
\newblock {Room-temperature spin injection from Fe into GaAs}.
\newblock {\em Phys. Rev. Lett.}, 87(1):016601, 2001.

\bibitem{alvarado1992observation}
S.~F. Alvarado and P.~Renaud.
\newblock {Observation of spin-polarized-electron tunneling from a ferromagnet
  into GaAs}.
\newblock {\em Phys. Rev. Lett.}, 68(9):1387, 1992.

\bibitem{hanbicki2002efficient}
A.~T. Hanbicki, B.~T. Jonker, G.~Itskos, G.~Kioseoglou, and A.~Petrou.
\newblock Efficient electrical spin injection from a magnetic metal/tunnel
  barrier contact into a semiconductor.
\newblock {\em Appl. Phys. Lett.}, 80(7):1240--1242, 2002.

\bibitem{fert2001conditions}
A.~Fert and H.~Jaffres.
\newblock Conditions for efficient spin injection from a ferromagnetic metal
  into a semiconductor.
\newblock {\em Phys. Rev. B}, 64(18):184420, 2001.

\bibitem{sau2010generic}
J.~D. Sau, R.~M. Lutchyn, S.~Tewari, and S.~D. Sarma.
\newblock Generic new platform for topological quantum computation using
  semiconductor heterostructures.
\newblock {\em Phys. Rev. Lett.}, 104(4):040502, 2010.

\bibitem{wei2016strong}
P.~Wei, S.~Lee, F.~Lemaitre, L.~Pinel, D.~Cutaia, W.~Cha, F.~Katmis, Y.~Zhu,
  D.~Heiman, and J.~Hone.
\newblock {Strong interfacial exchange field in the graphene/EuS
  heterostructure}.
\newblock {\em Nat. Mater.}, 15(7):711--716, 2016.

\bibitem{katmis2016high}
F.~Katmis, V.~Lauter, F.~S. Nogueira, B.~A. Assaf, M.~E. Jamer, P.~Wei,
  B.~Satpati, J.~W. Freeland, I.~Eremin, and D.~Heiman.
\newblock A high-temperature ferromagnetic topological insulating phase by
  proximity coupling.
\newblock {\em Nature}, 533(7604):513--516, 2016.

\bibitem{chen2017proximity}
H.~Q. Chen, B.~Li, and J.~L. Yang.
\newblock Proximity effect induced spin injection in phosphorene on magnetic
  insulator.
\newblock {\em ACS Appl. Mater. Interfaces}, 9(44):38999--39010, 2017.

\bibitem{liang2017magnetic}
X.~Liang, L.~J. Deng, F.~Huang, T.~T. Tang, C.~T. Wang, Y.~P. Zhu, J.~Qin,
  Y.~Zhang, B.~Peng, and L.~Bi.
\newblock {The magnetic proximity effect and electrical field tunable valley
  degeneracy in MoS$_2$/EuS van der Waals heterojunctions}.
\newblock {\em Nanoscale}, 9(27):9502--9509, 2017.

\bibitem{PhysRevLett.61.637}
J.~S. Moodera, X.~Hao, G.~A. Gibson, and R.~Meservey.
\newblock {Electron-Spin Polarization in Tunnel Junctions in Zero Applied Field
  with Ferromagnetic EuS Barriers}.
\newblock {\em Phys. Rev. Lett.}, 61:637--640, Aug 1988.

\bibitem{PhysRevLett.110.097001}
B.~Li, N.~Roschewsky, B.~A. Assaf, M.~Eich, M.~Epstein-Martin, D.~Heiman,
  M.~M\"unzenberg, and J.~S. Moodera.
\newblock Superconducting spin switch with infinite magnetoresistance induced
  by an internal exchange field.
\newblock {\em Phys. Rev. Lett.}, 110:097001, Feb 2013.

\bibitem{PhysRevLett.106.247001}
Y.~M. Xiong, S.~Stadler, P.~W. Adams, and G.~Catelani.
\newblock Spin-resolved tunneling studies of the exchange field in
  $\mathrm{EuS}/\mathrm{Al}$ bilayers.
\newblock {\em Phys. Rev. Lett.}, 106:247001, Jun 2011.

\bibitem{diesch2018creation}
S.~Diesch, P.~Machon, M.~Wolz, C.~S{\"u}rgers, D.~Beckmann, W.~Belzig, and
  E.~Scheer.
\newblock {Creation of equal-spin triplet superconductivity at the Al/EuS
  interface}.
\newblock {\em Nat. Commun.}, 9(1):1--8, 2018.

\bibitem{EuS-Pt}
J.~M. Gomez-Perez, X.~P. Zhang, F.~Calavalle, M.~Ilyn, C.~González-Orellana,
  M.~Gobbi, C.~Rogero, A.~Chuvilin, V.~N. Golovach, L.~E. Hueso, F.~S.
  Bergeret, and F.~Casanova.
\newblock Strong interfacial exchange field in a heavy metal/ferromagnetic
  insulator system determined by spin hall magnetoresistance.
\newblock {\em Nano Lett.}, 20(9):6815--6823, 2020.
\newblock PMID: 32786952.

\bibitem{PhysRevB.91.195310}
B.~A. Assaf, F.~Katmis, P.~Wei, C.~Z. Chang, B.~Satpati, J.~S. Moodera, and
  D.~Heiman.
\newblock Inducing magnetism onto the surface of a topological crystalline
  insulator.
\newblock {\em Phys. Rev. B}, 91:195310, May 2015.

\bibitem{PhysRevB.98.081403}
Q.~I. Yang and A.~Kapitulnik.
\newblock Two-stage proximity-induced gap opening in
  topological-insulator--insulating-ferromagnet
  ${({\mathrm{Bi}}_{x}{\mathrm{Sb}}_{1\ensuremath{-}x})}_{2}{\mathrm{te}}_{3}$--$\mathrm{EuS}$
  bilayers.
\newblock {\em Phys. Rev. B}, 98:081403, Aug 2018.

\bibitem{SnSe-EuS}
S.~Y. Yang, C.~Z. Wu, and N.~Marom.
\newblock {Topological properties of SnSe/EuS and SnTe/CaTe interfaces}.
\newblock {\em Phys. Rev. Materials}, 4:034203, Mar 2020.

\bibitem{liu2019coherent}
Y.~Liu, A.~Luchini, S.~Mart{\'\i}-S{\'a}nchez, C.~Koch, S.~Schuwalow, S.~A.
  Khan, T.~Stankevic, S.~Francoual, J.~R.~L. Mardegan, and J.~A. Krieger.
\newblock {Coherent Epitaxial Semiconductor--Ferromagnetic Insulator InAs/EuS
  Interfaces: Band Alignment and Magnetic Structure}.
\newblock {\em ACS Appl. Mater. Interfaces}, 12(7):8780--8787, 2020.

\bibitem{liu2020_strayAndExchange}
Y.~Liu, S.~Vaitiekėnas, S.~Martí-Sánchez, C.~Koch, S.~Hart, Z.~Cui,
  T.~Kanne, S.~A. Khan, R.~Tanta, S.~Upadhyay, M.~E. Cachaza, C.~M. Marcus,
  J.~Arbiol, K.~A. Moler, and P.~Krogstrup.
\newblock Semiconductor–ferromagnetic insulator–superconductor nanowires:
  Stray field and exchange field.
\newblock {\em Nano Lett.}, 20(1):456--462, 2020.
\newblock PMID: 31769993.

\bibitem{vaitiekenas2021zero}
S.~Vaitiek{\.e}nas, Y.~Liu, P.~Krogstrup, and C.~M. Marcus.
\newblock Zero-bias peaks at zero magnetic field in ferromagnetic hybrid
  nanowires.
\newblock {\em Nat. Phys.}, 17(1):43--47, 2021.

\bibitem{perdew1996generalized}
J.~P. Perdew, K.~Burke, and M.~Ernzerhof.
\newblock Generalized gradient approximation made simple.
\newblock {\em Phys. Rev. Lett.}, 77(18):3865, 1996.

\bibitem{yu2020machine}
M.~T. Yu, S.~Y. Yang, C.~Z. Wu, and N.~Marom.
\newblock {Machine learning the Hubbard U parameter in DFT+ U using Bayesian
  optimization}.
\newblock {\em NPJ Comput. Mater.}, 6(1):1--6, 2020.

\bibitem{ghosh2004electronic}
D.~B. Ghosh, M.~De, and S.~K. De.
\newblock Electronic structure and magneto-optical properties of magnetic
  semiconductors: Europium monochalcogenides.
\newblock {\em Phys. Rev. B}, 70(11):115211, 2004.

\bibitem{wachter1972optical}
P.~Wachter.
\newblock The optical electrical and magnetic properties of the europium
  chalcogenides and the rare earth pnictides.
\newblock {\em Crit. Rev. Solid State Mater. Sci.}, 3(2):189--241, 1972.

\bibitem{massidda1990structural}
S.~Massidda, A.~Continenza, A.~J. Freeman, T.~M. De~Pascale, F.~Meloni, and
  M.~Serra.
\newblock {Structural and electronic properties of narrow-band-gap
  semiconductors: InP, InAs, and InSb}.
\newblock {\em Phys. Rev. B}, 41(17):12079, 1990.

\bibitem{kim2009accurate}
Y.~S. Kim, K.~Hummer, and G.~Kresse.
\newblock {Accurate band structures and effective masses for InP, InAs, and
  InSb using hybrid functionals}.
\newblock {\em Phys. Rev. B}, 80(3):035203, 2009.

\bibitem{schlipf2013structural}
M.~Schlipf, M.~Betzinger, M.~Le{\v{z}}ai{\'c}, C.~Friedrich, and S.~Bl{\"u}gel.
\newblock {Structural, electronic, and magnetic properties of the europium
  chalcogenides: A hybrid-functional DFT study}.
\newblock {\em Phys. Rev. B}, 88(9):094433, 2013.

\bibitem{anisimov1991band}
V.~I. Anisimov, J.~Zaanen, and O.~K. Andersen.
\newblock {Band theory and Mott insulators: Hubbard U instead of Stoner I}.
\newblock {\em Phys. Rev. B}, 44(3):943, 1991.

\bibitem{dudarev1998electron}
S.~L. Dudarev, G.~A. Botton, S.~Y. Savrasov, C.~J. Humphreys, and A.~P. Sutton.
\newblock {Electron-energy-loss spectra and the structural stability of nickel
  oxide: An LSDA+ U study}.
\newblock {\em Phys. Rev. B}, 57(3):1505, 1998.

\bibitem{yang2020electronic}
S.~Y. Yang, N.~Schr{\"o}ter, S.~Schuwalow, M.~Rajpalk, K.~Ohtani,
  P.~KrogstrupGeorg, W.~Winkler, J.~Gukelberger, D.~Gresch, and G.~Aeppli.
\newblock {Electronic structure of InAs and InSb surfaces: density functional
  theory and angle-resolved photoemission spectroscopy}.
\newblock {\em arXiv preprint arXiv:2012.14935}, 2020.

\bibitem{yang2021first}
S.~Y. Yang, D.~Dardzinski, A.~Hwang, D.~I. Pikulin, G.~W. Winkler, and
  N.~Marom.
\newblock {First principles feasibility assessment of a topological insulator
  at the InAs/GaSb interface}.
\newblock {\em arXiv preprint arXiv:2101.07873}, 2021.

\bibitem{doi:10.1063/1.4858400}
R.~T. Tung.
\newblock The physics and chemistry of the schottky barrier height.
\newblock {\em Appl. Phys. Rev.}, 1(1):011304, 2014.

\bibitem{kresse1993ab}
G.~Kresse and J.~Hafner.
\newblock Ab initio molecular dynamics for liquid metals.
\newblock {\em Phys. Rev. B}, 47(1):558, 1993.

\bibitem{kresse1999ultrasoft}
G.~Kresse and D.~Joubert.
\newblock From ultrasoft pseudopotentials to the projector augmented-wave
  method.
\newblock {\em Phys. Rev. B}, 59(3):1758, 1999.

\bibitem{blochl1994projector}
P.~E. Bl{\"o}chl.
\newblock Projector augmented-wave method.
\newblock {\em Phys. Rev. B}, 50(24):17953, 1994.

\bibitem{steiner2016calculation}
S.~Steiner, S.~Khmelevskyi, M.~Marsmann, and G.~Kresse.
\newblock {Calculation of the magnetic anisotropy with projected-augmented-wave
  methodology and the case study of disordered $Fe_{1-x}Co_{x}$ alloys}.
\newblock {\em Phys. Rev. B}, 93(22):224425, 2016.

\bibitem{neugebauer1992adsorbate}
J.~Neugebauer and M.~Scheffler.
\newblock {Adsorbate-substrate and adsorbate-adsorbate interactions of Na and K
  adlayers on Al (111)}.
\newblock {\em Phys. Rev. B}, 46(24):16067, 1992.

\bibitem{tkatchenko2009accurate}
A.~Tkatchenko and M.~Scheffler.
\newblock Accurate molecular van der waals interactions from ground-state
  electron density and free-atom reference data.
\newblock {\em Phys. Rev. Lett.}, 102(7):073005, 2009.

\end{thebibliography}
\bibliographystyle{unsrt} %the RSC's .bst file
\end{document}